\documentclass[aps,prl,twocolumn,superscriptaddress]{revtex4}
\usepackage{graphicx}
\usepackage{enumerate}
\usepackage{hyperref}
\usepackage{textcomp}
\usepackage{amsmath}
\usepackage{amssymb}
\usepackage{pgf}
\usepackage{diagbox}
\usepackage{booktabs}
\usepackage{bbold}
\usepackage{color}
\usepackage{bbm}
\usepackage{soul}

\DeclareSymbolFont{matha}{OML}{txmi}{m}{it}
\DeclareMathSymbol{\varv}{\mathord}{matha}{118}

\def\ud{\mathrm{d}}
\def\ee{\mathrm{e}}

\def\bra#1{\langle#1 |}
\def\ket#1{| #1\rangle}

\newcommand{\nep}{\textrm{e}}

\newcommand{\QA}{\mathrm{\scriptscriptstyle QA}}

\newcommand{\btau}{\boldsymbol{\tau}}

\newcommand{\regular}{\mathrm{reg}}
\newcommand{\tcomp}{t_\mathrm{cc}}

\newcommand{\bvec}{\boldsymbol{v}}

\newcommand{\bbeta}{\boldsymbol{\beta}}
\newcommand{\bgamma}{\boldsymbol{\gamma}}

\newcommand{\n}{{\bf n}}

\newcommand{\calK}{{\mathcal{K}}}

\newcommand{\calP}{{\mathcal{P}}}

\newcommand{\opcdag}[1]{{\hat{c}^{\dagger}}_{#1}}
\newcommand{\opc}[1]{{\hat{c}^{\phantom \dagger}}_{#1}}

\newcommand{\PauliSigma}{\hat{\sigma}}
\newcommand{\PauliTau}{\hat{\tau}}
\newcommand{\bpaulitau}{{\widehat{\boldsymbol{\tau}}}}

\newcommand{\versorz}{{\hat{\boldsymbol{z}}}}

\newcommand{\versorb}{{\hat{\boldsymbol{b}}}}

\newcommand\norm[1]{\left\lVert#1\right\rVert}

\newcommand{\PBC}{{\scriptscriptstyle \mathrm{PBC}}}
\newcommand{\ABC}{{\scriptscriptstyle \mathrm{ABC}}}
\newcommand{\smallpm}{{\scriptscriptstyle (\pm)}}
\newcommand{\smallp}{{\scriptscriptstyle (+)}}
\newcommand{\smallm}{{\scriptscriptstyle (-)}}

\newcommand{\Texp}{\mathrm{T exp}}
\newcommand{\Tprod}[1]{\prod^{\leftarrow #1}}

\newcommand{\res}{\mathrm{res}}

\newcommand{\opt}{\mathrm{opt}}

\newcommand{\mgdigital}{\scriptscriptstyle\mathrm{digit}}

\newcommand{\Ham}{\widehat{H}}
\newcommand{\Uevol}{\widehat{U}}
\newcommand{\Rrot}{\mathcal{R}}

\newcommand{\eres}{\epsilon^{\res}}
\newcommand{\Tann}{\tau}
\newcommand{\Ptrot}{\mathrm{P}}

\newcommand{\Hred}{\widehat{\mathcal H}}
\newcommand{\Parity}{\calP}
\newcommand{\Parityop}{\widehat{\Parity}}
\newcommand{\Nred}{N_{\scriptscriptstyle \mathrm{R}}}
\newcommand{\target}{\scriptscriptstyle \mathrm{target}}

\graphicspath{{./}{../Figures/}{../Figures2}}

\begin{document}

\title{Optimal quantum control with digitized Quantum Annealing}

\author{Glen Bigan Mbeng}
\affiliation{SISSA, Via Bonomea 265, I-34136 Trieste, Italy}
\affiliation{INFN, Sezione di Trieste, I-34136 Trieste, Italy}
\author{Rosario Fazio}
\affiliation{Abdus Salam ICTP, Strada Costiera 11, 34151 Trieste, Italy}
\affiliation{Dipartimento di Fisica, Universit\`a di Napoli ``Federico II'', Monte S. Angelo, I-80126 Napoli, Italy}
%
\author{Giuseppe E. Santoro}
\affiliation{SISSA, Via Bonomea 265, I-34136 Trieste, Italy}
\affiliation{Abdus Salam ICTP, Strada Costiera 11, 34151 Trieste, Italy}
\affiliation{CNR-IOM Democritos National Simulation Center, Via Bonomea 265, I-34136 Trieste, Italy}

\begin{abstract}
We show how a digitized version of Quantum Annealing can be made optimal, realizing the best possible solution allowed
by quantum mechanics in the shortest time, without any prior knowledge on the location and properties of the spectral gap.
Our findings elucidate the intimate relation between digitized-QA, optimal Quantum Control, and recently proposed hybrid 
quantum-classical variational algorithms for quantum-state preparation and optimization.
We illustrate this on the simple benchmark problem of an unfrustrated antiferromagnetic Ising chain in a transverse field. 
\end{abstract}

\maketitle

{\em Introduction.---}Recent experimental advances in the world of quantum technologies have prompted the development of various 
quantum-based algorithms~\cite{Nielsen_Chuang:book}, some of which are suitable to run on available quantum devices, 
belonging to the class of Noisy Intermediate-Scale Quantum (NISQ) technologies~\cite{Preskill_Quantum2018}.
Two leading candidates in this area are Quantum Annealing \cite{Finnila_CPL94, Kadowaki_PRE98, Brooke_SCI99, Santoro_SCI02, 
Zecchina_PNAS2018}, and hybrid quantum-classical variational algorithms~\cite{Farhi_arXiv2014,Peruzzo_NatComm14,Zoller_NAT19}.

Quantum Annealing (QA)~\cite{Finnila_CPL94, Kadowaki_PRE98, Brooke_SCI99, Santoro_SCI02, Santoro_JPA06}, {\em alias} Adiabatic 
Quantum Computation~\cite{Farhi_SCI01, Albash_RevModPhys2018}, allows to solve hard optimization problems through a 
continuous-time adiabatic evolution of an appropriate quantum Hamiltonian. In this framework, the hardness of a problem is associated 
with the intrinsic difficulty in following the adiabatic ground state  when a (possibly exponentially) small spectral gap must be crossed to go 
from the initial state to the final target ground state ~\cite{Knysh_NatComm16, Zamponi_QA:review}. 
Different strategies have been proposed to cope with such a problem~\cite{Zeng_JPA2016, Zhuang_PRA14, Seoane_JPA2012, Zanardi_PRL2009}. 
Among them, the choice of the driving protocol is crucial for obtaining a quantum speed-up, see e.g.~\cite{Roland_PRA2002}. 
The schedule optimization, however, is believed to require, {\em in general}, information on the spectral gap for the problem,
posing a severe limitation to its implementation~\cite{Ambainis_arXiv2013, Wolf_Nat2015}.

Hybrid quantum-classical variational algorithms, instead, are insensitive to critical points and spectral information. 
They are based on classical minimization and invoke quantum digital processors to prepare a variational {\em Ansatz} for the 
problem~\cite{Farhi_arXiv2014, Peruzzo_NatComm14, Zoller_NAT19}. 
In the specific field of combinatorial optimization, this is accomplished by the Quantum Approximate Optimization Algorithm (QAOA) \cite{Farhi_arXiv2014} 
that operates through a depth-$\Ptrot$ circuit of digital unitary gates. 
In this framework, a problem is hard if it requires large-$\Ptrot$ (deep) quantum circuits to prepare a good {\em Ansatz}, or if the classical optimization landscape 
is complex and difficult to sample. 

Although QA and QAOA appear as unrelated models of computation, they are both computationally universal~\cite{Aharonov_QP04:preprint, 
Lidar_PRL07, Lloyd_arXiv2018}, suggesting that some connections might exist. 
Here we make a step forward in establishing this connection, by showing that one can construct an optimal QAOA solution which is {\em adiabatic}. 
Our contribution builds up on two recent interesting works. 
The first is the proposal for a fully {\em digitized} QA (dQA)~\cite{Martinis_Nat16} --- sharing technical similarities with the QAOA quantum 
circuit~\cite{Mbeng_DigitalQA_arXiv2019} ---, 
pointing towards a universal-gate approach to QA, with the bonus of the possibility of error-correction \cite{Kitaev_arXiv1998, Austin_PRA2012}. 
The second is the result of Yang {\em et al.}~\cite{Yang_PRX2017}, who showed that the digital nature of the QAOA {\em Ansatz} emerges naturally, when 
searching for an optimal protocol, from the ``bang-bang'' form predicted by the application of Pontryagin's principle~\cite{Dalessandro2007, Brif_NewJPhys2010}.
 
To demonstrate the construction of the optimal digitized-QA solution, we illustrate it for the benchmark problem of an Ising chain in a transverse field, 
where a detailed size-scaling analysis is feasible. 
We show that its QAOA depth-$\Ptrot$ quantum circuit admits in general many, $2^{\Ptrot}$, degenerate variational minima,  
all strictly having a residual energy error $\eres_{\Ptrot}=(2\Ptrot+2)^{-1}$.
We then show that among such $2^{\Ptrot}$ degenerate variational minima, 
there is a special {\em regular} optimal solution which can be constructed iteratively on $\Ptrot$, and which can be interpreted as an optimal digitized-QA schedule.
Such a schedule, obtained {\em without any spectral information}, 
is shown to be computationally optimal. The generality of such a procedure is finally discussed. 
It is worth stressing that when/if proved to be general, our procedure will allow to optimize Quantum Annealing protocols without any need to know the value of
the spectral gap. 

\begin{figure}
\centering
\includegraphics[width=75mm]{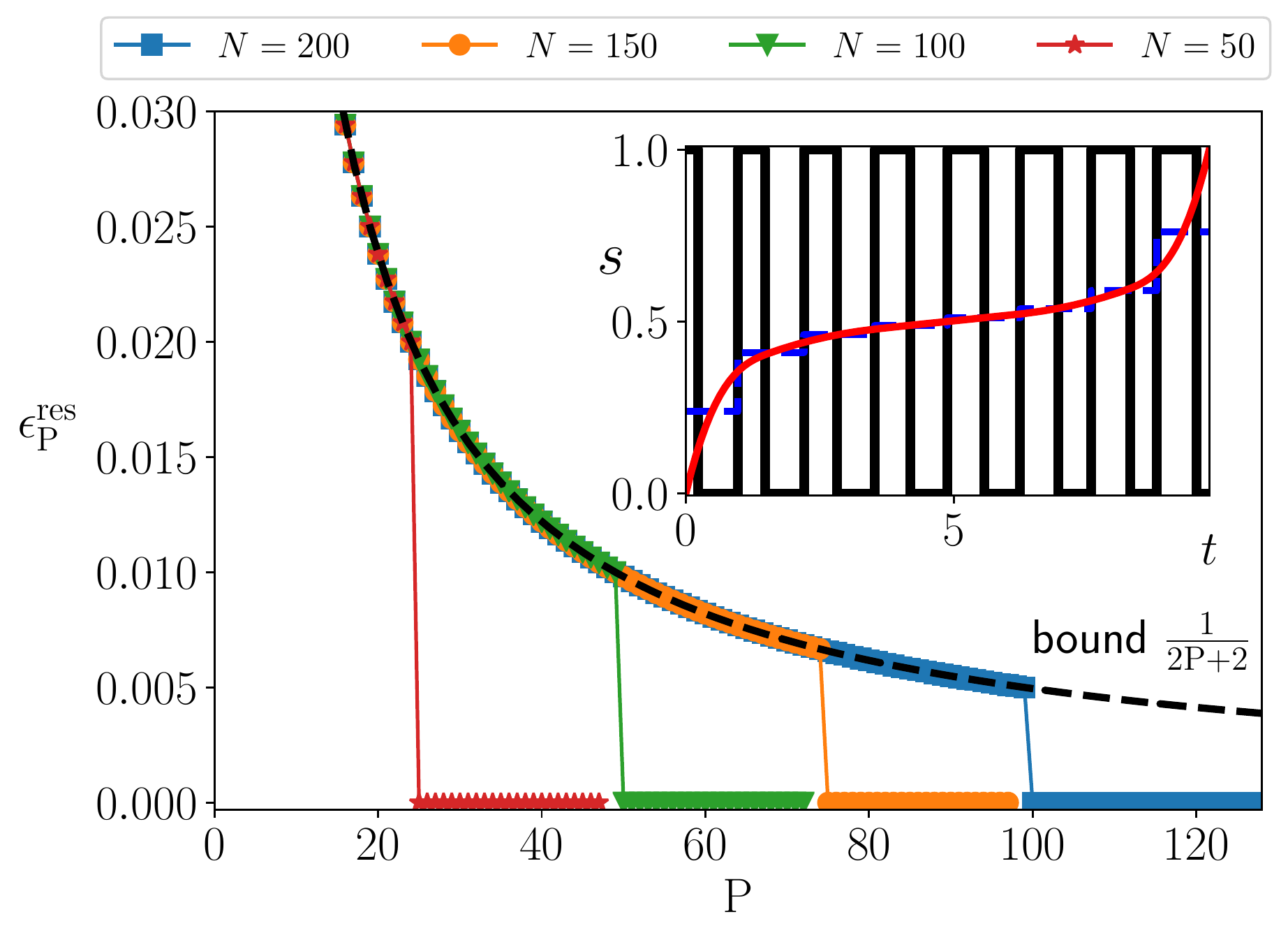}
\caption{Optimal residual energies $\eres_{\Ptrot}$ for the classical Ising $\Ham_z$ in Eq.~\eqref{eqn:Hz_classical} vs the number of steps $\Ptrot$, 
for various system sizes $N=50, 100, 150, 200$.  
	The symbols represent the data obtained by numerical minimization, the dashed line the theoretical bound in Eq.~\eqref{eqn:eres_final_main}. 
	The bound is saturated for $2\Ptrot<N$. For $2\Ptrot\geq N$, the residual energy drops to zero. 
	(Inset): Optimal schedule for $\Ptrot=8$ (corresponding to $\Tann=9.76$). Lines: solid red, the continuous-time $s(t)$; dashed blu, the step-discretization;
	solid black, the bang-bang digitized-QA.}
\label{fig:eres_bound_vs_P}
\end{figure}

{\em Methods.---}Consider the problem of finding the ground state of some spin-1/2 Hamiltonian $\Ham_{\target}$. 
In a continuous-time QA \cite{Kadowaki_PRE98,Santoro_JPA06,Albash_RevModPhys2018}, the target Hamiltonian $\Ham_{\target}$ is supplemented 
by a driving term, usually taken to be of the simple form  $\Ham_x=-\sum_{j=1}^N \PauliSigma^x_j$. 
For simplicity of presentation, we will assume that $\Ham_{\target}=\Ham_z + h \Ham_x$, where $\Ham_z$ contains only $\PauliSigma^z$ Pauli matrices.
In the simplest setting, one writes an interpolating QA Hamiltonian of the form: 
\begin{equation} 
\label{eqn:Hs}
		\Ham(s) = s \, \Ham_{\target} + (1-s) \, \Ham_x \;.
\end{equation}
The parameter $s$ is then varied in time, defining a schedule $s(t)$ interpolating from $s(0)=0$ to $s(\Tann)=1$, where
$\Tann$ is the total annealing time. 
Given any $s(t)$, and starting from the ground state of $\Ham_x$,  
$|\psi_0\rangle=
2^{-N/2}\left( |\!\uparrow\rangle + |\!\downarrow\rangle \right)^{\otimes N}$, 
the state of the system at time $\Tann$ is given by the Schr\"odinger evolution 
$|\psi(\Tann)\rangle = \Uevol_{\QA}(\Tann,0) | \psi_0 \rangle$ 
where $\Uevol_{\QA}(\Tann,0)$ is the evolution operator, formally expressed as a time-ordered exponential, 
$\Uevol_{\QA}(\Tann,0) = \Texp \left(-\frac{i}{\hbar} \int_0^\Tann \! \ud t' \, \Ham(s(t')) \right)$. 
%
%
By approximating the schedule $s(t)$ with a step function attaining $\Ptrot$ values $s_{m=1,\cdots, \Ptrot}\in (0,1]$, and corresponding 
evolution times $\Delta t_{m=1,\cdots, \Ptrot}$ such that $\sum_{m=1}^\Ptrot \Delta t_m = \Tann$ --- see sketch in Fig.~\ref{fig:eres_bound_vs_P}(inset) ---, 
and then taking a further Trotter splitting, the approximate evolution operator reads
\begin{equation} 
\label{eqn:Udigital}
		\Uevol_{\mgdigital} = \ee^{-i \beta_{\Ptrot} \Ham_x } \ee^{-i \gamma_{\Ptrot} \Ham_z} \cdots \ee^{-i \beta_1 \Ham_x } \ee^{-i \gamma_1 \Ham_z}\;, 
\end{equation}
where the parameters (for a lowest-order splitting $\gamma_m= s_m \Delta t_m/\hbar$ and $\beta_m=[(1-s_m)+hs_m] \Delta t_m/\hbar$) 
are such that:
\begin{equation} \label{eqn:sum_rule}
\sum_{m=1}^\Ptrot (\beta_m + (1-h) \gamma_m) = \frac{\Tann}{\hbar} \;.
\end{equation} 
Eq.~\eqref{eqn:Udigital} naturally leads to the QAOA~\cite{Farhi_arXiv2014}.
Indeed, one can regard the quantum state  
\begin{equation} \label{eqn:psi_qaoa}
|\psi_\Ptrot(\bgamma,\bbeta)\rangle  =
 \ee^{-i \beta_{\Ptrot} \Ham_x } \ee^{-i \gamma_{\Ptrot} \Ham_z} \cdots \ee^{-i \beta_1 \Ham_x } \ee^{-i \gamma_1 \Ham_z} | \psi_0\rangle \;,
\end{equation}
as variationally dependent on the $2\Ptrot$ parameters $(\bgamma,\bbeta)$.
Using a quantum device that prepares $|\psi_\Ptrot(\bgamma,\bbeta)\rangle$ and performs repeated measurements in the computational basis, 
we then evaluate the expectation value of the cost function Hamiltonian
\begin{equation}
E_\Ptrot(\bgamma,\bbeta) = \langle \psi_\Ptrot(\bgamma,\bbeta) |  \Ham_{\target} | \psi_\Ptrot(\bgamma,\bbeta) \rangle \;,  
\end{equation}
and minimize it through a classical algorithm.  
The global variational minimum $(\bgamma^{*},\bbeta^{*})$ determines a correspondingly optimal state 
$\ket{\psi_\Ptrot(\bgamma^*, \bbeta^*)}$, whose energy $E_\Ptrot^{\opt}=E_\Ptrot(\bgamma^{*},\bbeta^{*})$ is, by construction, 
a monotonically decreasing function of $\Ptrot$.
%
%
%
Remarkably, this QAOA approach is computationally universal~\cite{Lloyd_arXiv2018}, although this does not guarantee efficiency or speed-up~\cite{Hastings_arXiv2019}. 
Interestingly, Yang \textit{et al.}~\cite{Yang_PRX2017} have shown that such a protocol naturally emerges as an application of the
Pontryagin's principle \cite{Dalessandro2007, Brif_NewJPhys2010} of optimal control: restricting $s(t)$ in the interval $[0,1]$ leads to
optimal {\em bang-bang} schedules, i.e., $s(t)$ having a square-wave form between the two extremal values $1$ and $0$,
see inset of Fig.~\ref{fig:eres_bound_vs_P}. 


We can quantify the degree to which a variational QAOA state $\ket{\psi_\Ptrot(\bgamma,\bbeta)}$ 
approximates the solution of the quantum problem with the rescaled residual energy \cite{Santoro_SCI02}
\begin{equation} \label{eqn:e_res} 
    \eres_{\Ptrot}(\bgamma, \bbeta) =  \frac{E_\Ptrot(\bgamma, \bbeta)-E_{\min}}{E_{\max}-E_{\min}}  \;,
\end{equation}
where $E_{\min}$ and $E_{\max}$ are the minimum and maximum eigenvalues of $\Ham_{\target}$. 
$\eres_{\Ptrot}$ is normalized in such a way that $\eres_\Ptrot \in [0,1]$ and that 
$\eres_\Ptrot(\bgamma, \bbeta)=0$ if and only if $\ket{\psi_\Ptrot(\bgamma, \bbeta)}$ is a ground state of the $\Ham_{\target}$. 

{\em A variational bound.---} Consider for illustration the simple problem of finding the classical ground state of the Ising antiferromagnetic chain
\begin{equation} 
\label{eqn:Hz_classical}
		\Ham_{\target} = \Ham_z = \sum_{j=1}^N \PauliSigma^z_j \PauliSigma^z_{j+1} \;,
\end{equation}
where $\PauliSigma^z_j$ is a Pauli matrix at site $j$, and periodic boundary conditions (PBC) are assumed.  
Here $E_{\min}=-N$ and $E_{\max}=N$. 
One can show~\cite{Mbeng_HybridQVS_arXiv2019} that for the Ising chain problem:
\begin{equation} \label{eqn:eres_final_main}
\eres_{\Ptrot} \ge 
\left\{ \begin{array}{ll} \displaystyle \frac{1}{2 \Ptrot + 2} & \mbox{for} \hspace{2mm} 2\Ptrot < N \vspace{3mm} \\
0  & \mbox{for} \hspace{2mm} 2\Ptrot \ge N
\end{array} \right. \;.
\end{equation}
A general derivation of the bound in Eq.~\eqref{eqn:eres_final_main} relies on the locality and translational invariance of the problem,
as reported elsewhere~\cite{Mbeng_HybridQVS_arXiv2019}. 
The gist of the proof is that for $2\Ptrot<N$ we can effectively deal with a reduced spin chain of length $\Nred=2\Ptrot+2$, and an extra freedom on the
boundary condition: For $2\Ptrot\ge N$ we set $\Nred=N$ and use PBC, while for $2\Ptrot<N$ we can use {\em anti-periodic} boundary conditions (ABC), 
leading to Eq.~\eqref{eqn:eres_final_main}.
%
Here we proceed by using a Jordan-Wigner transformation~\cite{JordanWigner_ZPhys1928,Wang_PRA2018}, such that the relevant Hamiltonians 
can be expressed as a sum of independent two-level systems labeled by a wave-vector $k$ whose values depend on the boundary conditions used. 
In particular, we define $\calK_{\PBC}=\{ \frac{\pi}{\Nred}, \frac{3 \pi}{\Nred} \cdots,\frac{ (\Nred-1)\pi }{\Nred} \}$ and 
$\calK_{\ABC}=\{ \frac{2\pi}{\Nred}, \frac{4\pi}{\Nred}, \cdots, \frac{(\Nred-2)\pi}{\Nred} \}$ to be the set of $k$-vectors associated to
PBC and ABC, respectively, for the spin chain. 
%
The final result (see Supplementary Information) 
is that the residual energy can be decomposed, for $2\Ptrot < N$, as:
\begin{equation}  \label{eqn:eres_abc}
\eres_{\Ptrot}(\bgamma, \bbeta) \stackrel{\scriptscriptstyle 2\Ptrot < N}{=}  
\frac{1}{2\Ptrot+2} + \frac{1}{2\Ptrot+2}{\displaystyle \sum_{k}^{\calK_{\ABC}}} \epsilon_{k}(\bgamma, \bbeta) \;,
\end{equation}
while for $2\Ptrot\ge N$ we get:
\begin{equation}  \label{eqn:eres_pbc}
\eres_{\Ptrot}(\bgamma, \bbeta) \stackrel{\scriptscriptstyle 2\Ptrot\ge  N}{=}  \frac{1}{N}{\displaystyle \sum_{k}^{\calK_{\PBC}}} \epsilon_{k}(\bgamma, \bbeta) \;.
\end{equation}
The function $\epsilon_{k}(\bgamma, \bbeta)\ge 0$, given by
\begin{equation} \label{eqn:eresk_geometrical_def}
\epsilon_{k}(\bgamma, \bbeta)= 1-  \versorb_k^T \left( \Tprod{\Ptrot}_{m=1}  \Rrot_{\versorz}(4 \beta_m )\Rrot_{\versorb_k}(4 \gamma_m ) \right) 
\versorz \;,
\end{equation}
is expressed in terms of the scalar product of the unit vector $\versorb_{k}=(-\sin k,0, \cos k)^T$ 
with the unit vector 
obtained by applying to $\versorz=(0,0,1)^T$ the $2\Ptrot$ successive $3\times 3$ rotation matrices $\Rrot_{\hat{\n}}(\theta)$,
around the axis $\hat{\n}=\versorz$ and $\hat{\n}=\versorb_{k}$ by rotation angles $4\beta_m$ and $4 \gamma_m$, respectively.  
(Here $\Tprod{\Ptrot}_{m=1}$ denotes a ``time-ordered'' product, where $m$ increases from right to left.)
%
It assumes its minimum value when $\epsilon_{k}(\bgamma, \bbeta)=0$.
%
%
%
These can be regarded as a set of non-linear constraints, whose number depends on the number of $k$-vectors involved, hence on the boundary conditions.  
The minimum residual energy is obtained by simultaneously minimizing all the addends of the $k$-sum in Eqs.~(\ref{eqn:eres_abc}-\ref{eqn:eres_pbc}). 
Notice that:
%
\begin{description}
    \item[For $\mathbf{2P<N}$] the number of constraints is $2|\calK_{\ABC}|=2\Ptrot$ 
    --- corresponding to $|\calK_{\ABC}|=\Ptrot$ constraints for $3$-dimensional unit vectors ---, hence equal to the number of variables $2\Ptrot$.
    The equations $\epsilon_{k}(\bgamma, \bbeta)=0$ have a finite set of discrete solutions. 
    When all equations are satisfied we get, see \eqref{eqn:eres_abc}, the optimal $\eres(\bgamma^*,\bbeta^*)=(2\Ptrot+2)^{-1}$.
    \item[For $\mathbf{2P\ge N}$] the number of variables  $2\Ptrot$ is equal or exceeds the number of constraints 
    $2|\calK_{\PBC}|=N$. The equations $\epsilon_{k}(\bgamma, \bbeta)=0$ have discrete solutions (for $2\Ptrot=N$) or a continuum of solutions 
    (for $2\Ptrot>N$) where,  see \eqref{eqn:eres_pbc}, $\eres(\bgamma^*,\bbeta^*)=0$. 
\end{description}

Figure~\ref{fig:eres_bound_vs_P} illustrates the minimum residual energy obtained numerically 
---
with the Broyden-Fletcher-Goldfard-Shanno (BFGS) algorithm~\cite{Nocedal_book2006}, 
using back-propagation to compute the required gradients ---
for different values of $N$, as a function of $\Ptrot$. 
%
When $2\Ptrot\ge N$, $\eres_{\Ptrot}$ drops to $0$, as predicted by the counting argument given above. 

\begin{figure}
\centering
\includegraphics[width=75mm]{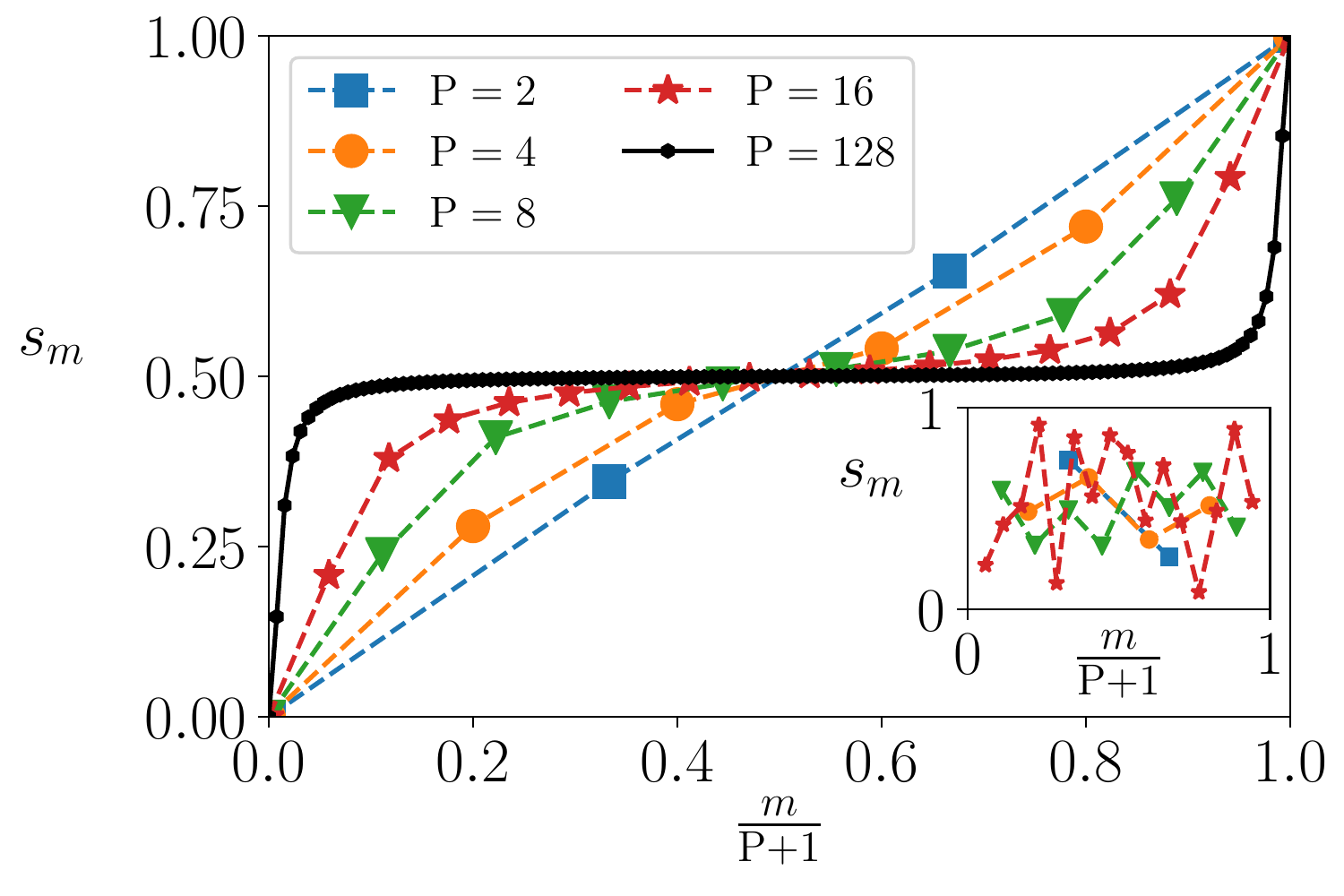}
\caption{Iterative construction of the ``regular'' solution for targeting $\Ham_z$ in Eq.~\eqref{eqn:Hz_classical} for increasing $\Ptrot$. 
    Here $s_m = \gamma_m/(\gamma_m + \beta_m)$. 
    The inset shows a set of generic optimal solutions obtained by initializing $(\bgamma,\bbeta)$ randomly, producing irregular solutions. 
}
\label{fig:regular_schedule}
\end{figure}

{\em The optimal digitized-QA solution.---}
For $2\Ptrot<N$, $\Nred=2\Ptrot+2$ and the QAOA landscape is independent of the system size $N$, see \eqref{eqn:eres_abc}. 
We find (numerically) that there are $2^{\Ptrot}$ {\em degenerate} minima all sharing the same $\eres_{\Ptrot} = (2\Ptrot+2)^{-1}$,  
corresponding to equivalent optimal choices for the $\gamma_m$ and $\beta_m$, most of which lack any structure or pattern. 
Here we show how to construct a regular schedule with a smooth $\Ptrot\! \to \!\infty$ limit, which is digitally adiabatic. 
We proceed iteratively in $\Ptrot$~\cite{Lukin_arXiv2018}: 
the optimal solution at level $\Ptrot$ is obtained by using, as an initial guess for $(\bgamma,\bbeta)$, the regular solution obtained for $\Ptrot'<\Ptrot$. 
Figure~\ref{fig:regular_schedule} illustrates this procedure.
For $\Ptrot=2$ the solution, in terms of $s_m=\gamma_m/(\gamma_m+\beta_m)$, 
nearly coincides with the expected ``linear-schedule'' $s(t)=t/\Tann$, a standard choice in 
continuous-time QA~\cite{Kadowaki_PRE98, Santoro_JPA06, Albash_RevModPhys2018}, which is used here as the starting point in searching for the minimum.
We next consider $\Ptrot=4$ and start the minimization search from the interpolation of the $\Ptrot=2$ values. 
The minimum found now deviates from the linear interpolation. 
Proceeding further, with $\Ptrot=8, 16, \cdots$, we get the solutions shown in Fig.~\ref{fig:regular_schedule}, 
whose inset, by contrast, illustrates the ``irregular'' values of $s_m$ obtained by starting the search from a random initial point. 
As said, there are $2^{\Ptrot}$ {\em degenerate global minima} all sharing the same $\eres_{\Ptrot} = (2\Ptrot+2)^{-1}$ for $2\Ptrot<N$: 
the minimum found by the BFGS routine depends on the choice of the initial guess for $(\bgamma, \bbeta)$.
Summarizing, among the vast majority of irregular solutions, one can single-out, through an appropriate iterative
search scheme, a {\em regular} solution whose parameters $s_m$ appear to have a well recognizable ``structure'', which is
found to have a data collapse (not shown) to a simple scaling form 
$s_{\Tann}(t) = \frac{1}{2} + \frac{1}{\Tann} f\left( \frac{t}{\Tann} \right)$. 
%
%
We have verified that the regular solution $(\bgamma^{\regular},\bbeta^{\regular})$ indeed defines an {\em adiabatic} discrete-time digital 
dynamics~\cite{Dranov_JMatPhys1998}, hence realizing an optimal digitized-QA schedule~\cite{Martinis_Nat16}.
%

\begin{figure}
\centering
    \includegraphics[width=75mm]{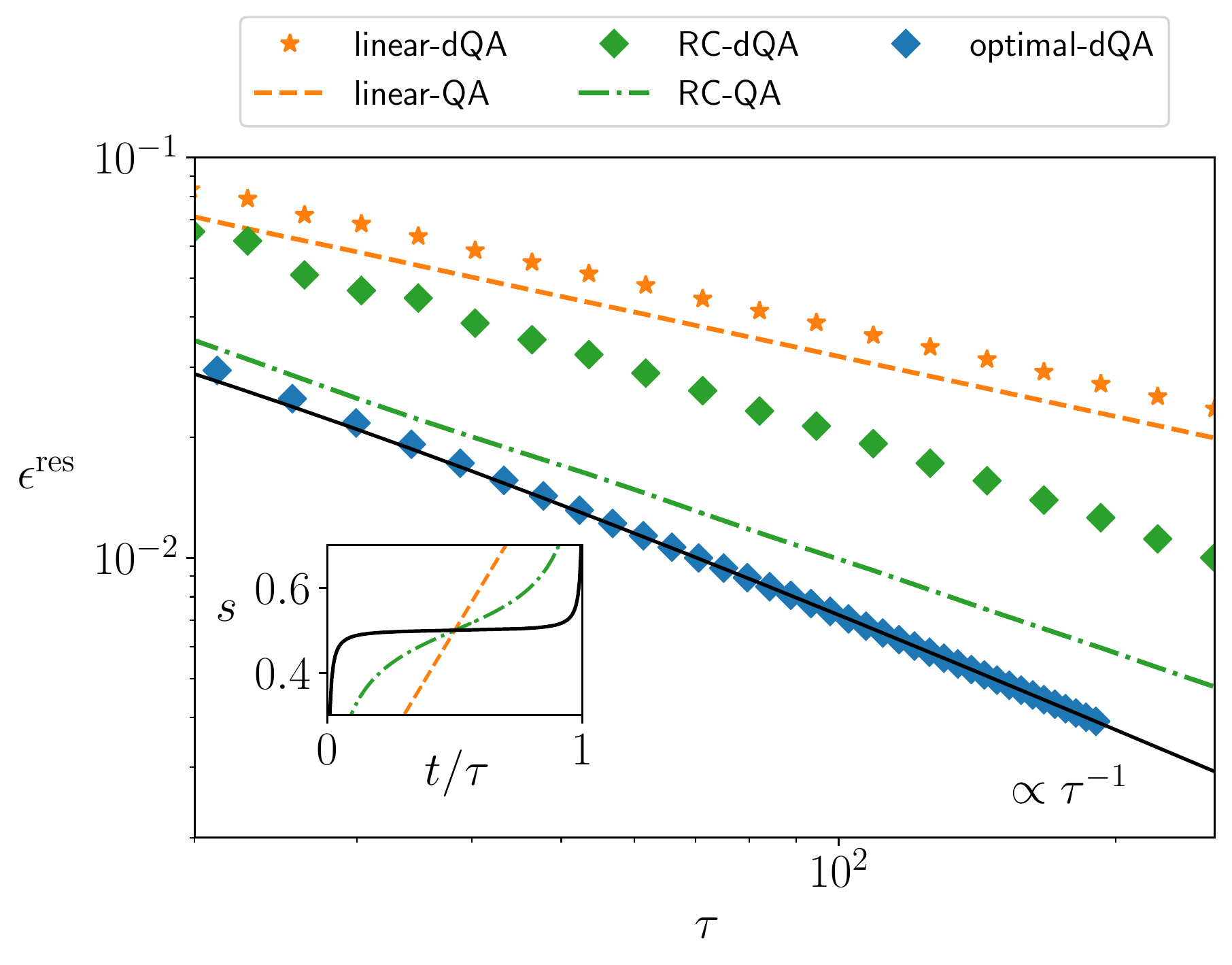}
    \caption{Scaling of the residual energy for various QA schedules in the infinite Ising chain problem. 
    All digitized-QA (dQA) data assume $\Delta t_m=1$ (in units of $\hbar/J$).  
    The linear QA/dQA (orange dashed line/stars) show a Kibble-Zurek exponent $\eres\sim \Tann^{-1/2}$. 
    The optimized Roland-Cerf \cite{Roland_PRA2002} QA/dQA (green dash-dot line/diamonds, RC) 
    shows $\eres\sim \Tann^{-\alpha}$ with $\alpha\approx 0.75$. 
    The blue hexagons represent the regular optimal dQA results, the solid line being the best fit with $1/(a\Tann+b)$.
    The inset shows $s(t)$ at fixed $\tau=32$ for the different schedules.}
    \label{fig:QAOA_ndef}
\end{figure}

{\em Speed-up and generalizations.---}One might ask how such optimal digitized-QA solution compares with other standard QA approaches for the ordered Ising chain.
Specifically, one standard route is that of a linear-schedule continuous-time QA, henceforth referred to as ``linear-QA'', where $s(t)=t/\Tann$. 
This is well known~\cite{Dziarmaga_PRL05,Zurek_PRL05} to lead to a power-law scaling~\cite{kibble76, zurek85, Polkovnikov_RMP11} of the
residual energy $\eres(\Tann) \sim \Tann^{-1/2}$. 
Figure \ref{fig:QAOA_ndef} shows that both linear-QA and linear-dQA (obtained by digitalization with $\Delta t_m=1$, in units of $\hbar/J$)
display the correct KZ behaviour $\eres(\Tann) \sim \Tann^{-1/2}$.
%
Next, we consider an optimized Roland-Cerf schedule \cite{Roland_PRA2002} --- similar results (not shown) are obtained for the power-law schedule proposed
in Ref.~\onlinecite{Polkovnikov_PRL2008} --- which exploits knowing the location of the critical point $s_c=1/2$.
%
%
%
Optimizing the schedule parameters numerically, it produces an improvement over linear-QA, with $\eres \sim \Tann^{-\alpha}$, where $\alpha\sim 0.75$.
%
Finally, Fig.~\ref{fig:QAOA_ndef} shows the residual energy corresponding to the optimal digitized-QA solution, with $\tau$ calculated from \eqref{eqn:sum_rule}. 
Here the behaviour of $\eres(\Tann)$ shows the optimal power-law $\eres\sim \Tann^{-1}$, consistently with the bound $\eres_{\Ptrot}\ge (2\Ptrot+2)^{-1}$
and with $\tau \propto \Ptrot$. 

The regular optimal dQA solution has the best possible performance, saturating the residual energy bound: $\eres\sim \Tann^{-1}$. 
However, such a quadratic speed-up over the plain KZ exponent comes with an extra cost to find the global QAOA minimum. 
How would $\eres$ decrease as a function of the total computational cost $\tcomp$?
Suppose we agree that the cost associated to the ``quantum oracle'' estimation of a length-$\Ptrot$ circuit scales with $\Ptrot$, 
the number of unitaries in $\ket{\psi_\Ptrot(\bgamma,\bbeta)}$, so that 
running the algorithm $n_{\rm iter}$ times to find $(\bgamma^\opt,\bbeta^{\opt})$
gives $\tcomp \propto n_{\rm iter} \Ptrot$.  
We find (not shown) that $n_{\rm iter}^{\rm random} \propto \Ptrot^2$ for a search starting from a random initial point, while  
$n_{\rm iter}^{\rm regular} \propto \Ptrot^{1/2}$ for the iterative search of the regular solution. 
Hence, the quadratic speed-up is wasted in the random case, 
$\eres_{\rm random} \sim \Ptrot^{-1} \sim \tcomp^{-1/3}$ -- worse than KZ $\sim \tcomp^{-1/2}$ ---,
while a speed-up survives when the regular optimal solution is constructed:
$\eres_{\rm  regular} \sim \Ptrot^{-1} \sim \tcomp^{-2/3}$.
%
%
To improve over linear-dQA, one {\em must} use a recursive initialization, leading to an optimal-dQA. 

\begin{figure}
\centering
\includegraphics[width=75mm]{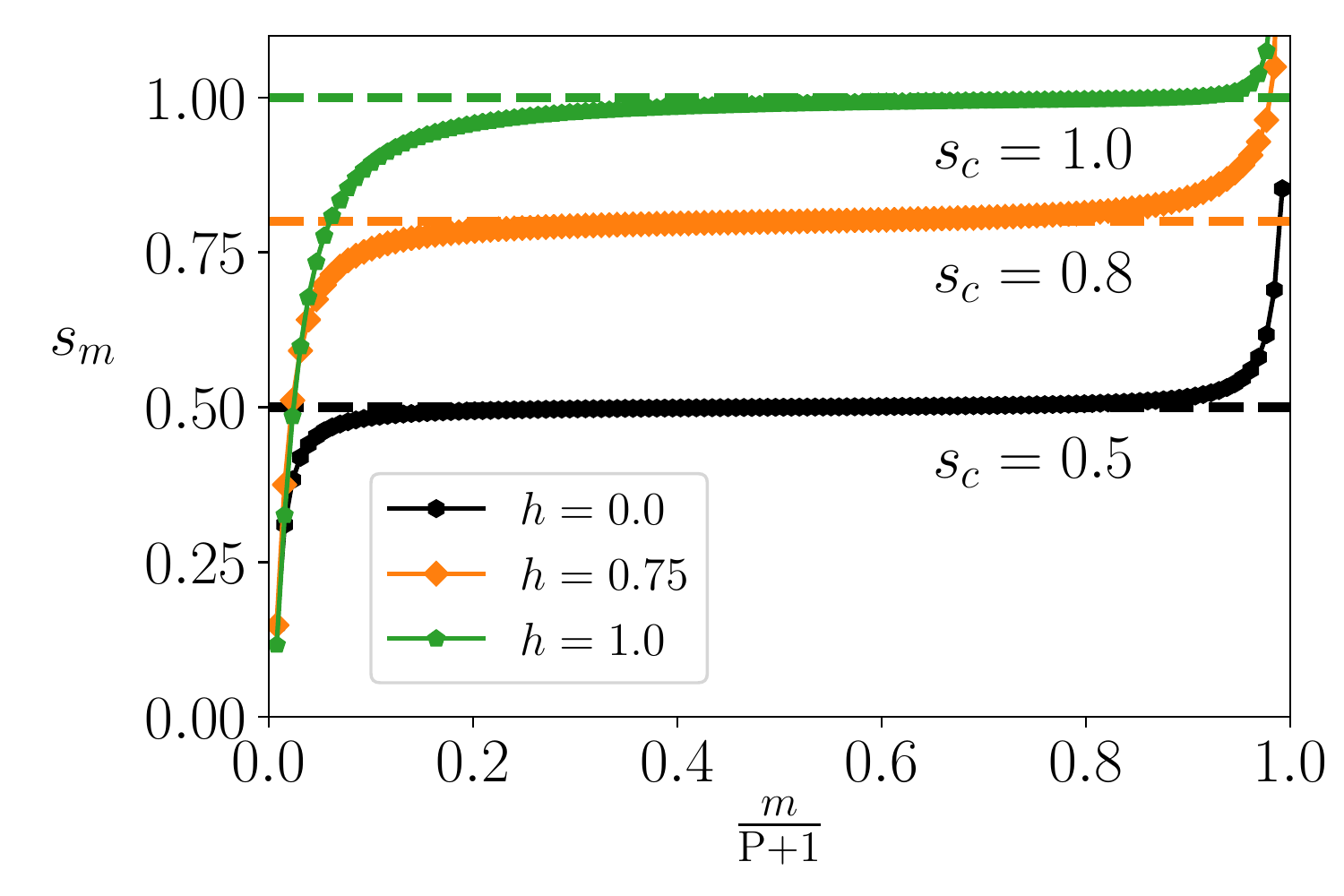}
\caption{Optimal solutions for $\Ptrot=128$ when targeting the Ising ground state of $\Ham_{\target}=\Ham_z+h\Ham_x$ at different $h$.
}
\label{fig:regular_schedule_various-h}
\end{figure}

Figure~\ref{fig:regular_schedule_various-h} shows that the procedure we introduced leads to smooth $s(t)$ schedules also when
the target Hamiltonian is the transverse-field Ising model at a non-zero field $h$, $\Ham_{\target}=\Ham_z+h\Ham_x$. 
Remarkably, $s(t)$ appears to flatten close to the critical point $s_c=2/(1-h)$, but this is achieved automatically by the 
iterative optimization protocol described, without any prior spectral information.
We have verified that the same procedure is successful in targeting the ground state of a general anisotropic XY-model chain, 
as reported elsewhere \cite{Glen_lungo}.  
%

{\em Conclusions.---}We unveiled deep connections between Quantum Annealing (QA), in its digitized version~\cite{Martinis_Nat16},
with the hybrid quantum-classical variational approach known as QAOA~\cite{Farhi_arXiv2014}, 
realizing optimal control of the schedule without spectral information. 

Generally speaking, the question of how and when an adiabatic optimal solution can be constructed
is an issue that deserves further investigations. 
The ingredients that must be carefully considered are the locality of the Hamiltonian, and whether the critical point separating
the final target state from the initial one has a finite-size gap that closes as a power-law or, rather, exponentially fast with increasing system-size $N$.

Another interesting issue has to do with the role of disorder \cite{Caneva_PRB2007, Mbeng_PRB2019}. 
We have verified, and will report elsewhere, that the perfect degeneracy of the optimal solutions found in the present translationally invariant case is 
broken in the presence of disorder: the variational energy landscape becomes rugged, and the search for the global optimal solution turns 
to be computationally harder.   
Further scrutiny is needed to investigate the quality of the adiabatic regular solution in a situation in which a large number of non-degenerate minima is present. 
The application of Machine Learning ideas \cite{Bukov_PRX18, Troyer_PRA16, Crooks_arXiv2018} to such complex minimization problems appears 
to be a fascinating perspective. 
      
We acknowledge fruitful discussions with L. Arceci and M. Wauters. 
Research was partly supported by EU H2020 under ERC-ULTRADISS, Grant Agreement No. 834402.
RF and GES acknowledge that their research has been conducted within the framework of the Trieste Institute for Theoretical Quantum Technologies (TQT).

\bibliography{../../BiblioQIC,../../BiblioQAOA,../../BiblioQA,../../BiblioNishimori,../../BiblioQIsing} 

\begin{widetext}
\section{Supplemental Material}

This Supplementary Information contains useful material related to the Jordan-Wigner transformation. 
This tool is used to diagonalize the quantum Ising chain and to compute the residual energy associated with QAOA variational states $\ket{\psi_{\Ptrot}(\bgamma,\bbeta)}$.

The Jordan-Wigner (JW) transformation for the quantum Ising chain is rather standard~\cite{JordanWigner_ZPhys1928, Lieb_AnnPhys1961}.   
For the continuous-time QA and QAOA with PBC, see for instance Refs.~\cite{Dziarmaga_PRL2005,Wang_PRA2018}. 
In our approach however (see also Ref.~\cite{Mbeng_HybridQVS_arXiv2019}) the boundary conditions play a crucial role. 
In particular, a non-trivial variational bound~\cite{Mbeng_HybridQVS_arXiv2019} was obtained by considering a reduced spin chain with 
anti-periodic boundary conditions (ABC), rather than PBC. 
We therefore present here a brief unified derivation, valid for both PBC ($+$) and ABC ($-$), 
discussing the application of JW to the digital dynamics (digitized-QA or QAOA) of a reduced quantum Ising chain of $\Nred$ spins. 
Specifically, starting from the initial state 
$|\psi_0\rangle=|+\rangle^{\otimes N}$, with $|+\rangle =\left( |\!\uparrow\rangle + |\!\downarrow\rangle \right)/\sqrt{2}$, 
we will consider the digitized dynamics obtained by alternating the following reduced Hamiltonians~\cite{Mbeng_HybridQVS_arXiv2019}:
\begin{eqnarray} 
\Hred^{\smallpm}_x &=& -\sum_{j=1}^{\Nred} \PauliSigma^x_j \label{eqn:Hred_x_abc/pbc}\\
\Hred^{\smallpm}_z &=& \sum_{j=1}^{\Nred-1}\PauliSigma^z_j \PauliSigma^z_{j+1} 
\pm \, \PauliSigma^z_{\Nred} \PauliSigma^z_{1} \;. \label{eqn:Hred_z_abc/pbc}
\end{eqnarray}

\subsection{Jordan-Wigner transformation for Hamiltonian diagonalization}
The global parity $\Parityop=\prod_{n=1}^{\Nred}\PauliSigma^x_n$ is a conserved quantity for all the Hamiltonians we consider. 
Therefore, since the initial state has an even parity $\Parityop\ket{\psi_0}=\ket{\psi_0}$, we can restrict our analysis to such a subspace.
When restricted to the even parity sector, a Jordan-Wigner transformation~\cite{JordanWigner_ZPhys1928},
$\PauliSigma^x_j = 1- 2\opcdag{j} \opc{j}$, 
$\PauliSigma^z_j = -(\opc{j} +\opcdag{j})\exp \left(-i\pi\sum_{l=1}^{j-1}\opcdag{l} \opc{l}\right)$, 
%
%
maps the spin system to free spinless fermions on a lattice, where $\opcdag{j}$ and $\opc{j}$ respectively 
create and annihilate a fermion at site $j$. 
After this transformation the Hamiltonians take the form 
\begin{eqnarray}
\Hred^{\smallpm}_x&=& \sum_{j=1}^{\Nred - 1}  (\opcdag{j}\opc{j} 
- \opc{j}\opcdag{j}) \\
\Hred^{\smallpm}_z&=&\sum_{j=1}^{\Nred-1}(\opcdag{j} - \opc{j})(\opcdag{j+1}+\opc{j+1})
\mp (\opcdag{\Nred} - \opc{\Nred})(\opcdag{1}+ \opc{1})\,,\\
\end{eqnarray}
where PBC for the spins are mapped into ABC for the fermions, and vice-versa.
A Fourier transformation can then be used to decompose the system into a set of decoupled two-level systems. 
This is done by introducing a set of wave-vectors $\tilde{\calK}^{\smallpm}$ that, to be consistent with the boundary conditions, must be taken to be
\begin{eqnarray}
\mathrm{spin-PBC}: \,\,\tilde{\calK}^\smallp &=& 
\left\{\pm \pi\frac{2n-1}{\Nred} \mbox{ for } n=1, 2,\cdots,\frac{\Nred}{2}\right\} \label{eqn:Ktilde_pbc} \\
\mathrm{spin-ABC}: \,\, \tilde{\calK}^\smallm &=& 
\left\{ \pm 2\pi\frac{n}{\Nred} \mbox{ for } n=1, 2,\cdots,\frac{\Nred}{2}-1\right\}
\cup\bigg\{ 0,\pi \bigg\} \label{eqn:Ktilde_abc}
\end{eqnarray}
and substituting
\begin{eqnarray}
\opcdag{j} &=& \frac{\ee^{i\pi/4}}{\sqrt{\Nred}}\sum_{k}^{\tilde{\calK}^{\smallpm}} \ee^{+i k j}\, \opcdag{k}\,,
\end{eqnarray}
where $\opcdag{k}$ creates a fermion with wave-vector $k$, and the appropriate set $\tilde{\calK}^{\smallpm}$ is assumed to be used in the sum over $k$. 
In terms of these Fourier modes the Hamiltonians 
%
%
decompose into pairs of modes with opposite momenta $k$ and $-k$. 
The main difference between PBC and ABC emerges at this level. 
Indeed, the special modes with $k=0,\pi$, which appear only with spin-ABC, are self-conjugate and 
do not couple to any other mode. 
A direct consequence of this is that, with spin-ABC, the number operators associated with such modes are conserved quantities. 
In particular, since these modes are absent in the initial state 
$\opcdag{0}\opc{0} \ket{\psi_0}=\opcdag{\pi}\opc{\pi} \ket{\psi_0}=0$, 
we can restrict ourselves to the subspace where the $k=0, \pi$ modes are absent. 
The Hamiltonians then read
\begin{eqnarray}\label{eqn:Hredx_relevant_k}
\Hred^{\smallpm}_x&=& (\pm 1-1) + 2\sum_{k}^{ \calK^{\smallpm}}  \left( \opcdag{k}\opc{k} - \opc{-k}\opcdag{-k} \right) \\
\Hred^{\smallpm}_z&=& 2 \sum_{k}^{ \calK^{\smallpm}} \left[ \sin{k}\, \left(\opcdag{k}\opcdag{-k} + \opc{-k}\opc{k}\right) +
\cos k  \, \left( \opcdag{k}\opc{k}-\opcdag{-k}\opc{-k} \right) \right] \;, \label{eqn:Hredz_relevant_k}
\end{eqnarray}
where the sum over $k$ now runs over the appropriate set of dynamically active (and positive) wave-vectors 
$\calK$ given by:
\begin{eqnarray}
\mathrm{spin-PBC}: \,\,\calK^\smallp &=& 
\left\{\frac{(2n-1)\pi}{\Nred} \mbox{ for } n=1, 2,\cdots,\frac{\Nred}{2}\right\}\label{eqn:K_pbc} \\
\mathrm{spin-ABC}: \,\, \calK^\smallm &=& 
\left\{ \frac{2n\pi}{\Nred} \mbox{ for } n=1,2,\cdots,\frac{\Nred}{2}-1\right\} \;. \label{eqn:K_abc}
\end{eqnarray}
A further inspection of Eqs.~\eqref{eqn:Hredx_relevant_k} and ~\eqref{eqn:Hredz_relevant_k} 
also reveals that each pairs' parity operator 
$\Parityop_k=\ee^{i\pi (\opcdag{k}\opc{k}+\opcdag{-k}\opc{-k})}$ is conserved. 
Again, since $\Parityop_k\ket{\psi_0}=1$ for all $k\in \calK^{\smallpm}$, we can restrict our analysis to the even-parity subspace 
$\Parity_k=1$ for all $k\in \calK^{\smallpm}$. 

Finally, in this even-parity subspace, the system is equivalent to a collection of decoupled two-level systems (or pseudo-spins),  
for instance through the identification $\ket{\!\uparrow_k}=\ket{0}$ and $\ket{\!\downarrow_k}=\opcdag{k}\opcdag{-k}\ket{0}$. 
The number of independent pseudo-spins for spin-PBC is $\left| \calK^\smallp \right|=\Nred/2$ while for 
spin-ABC, due to the absence of the $k=0,\pi$ modes, it is given by $\left|\calK^\smallm \right|=\Nred/2-1$.  
By introducing the pseudo-spin Pauli operators $\bpaulitau_k = (\PauliTau^x_k, \PauliTau^y_k, \PauliTau_k^z)^T$, 
the Hamiltonians read
\begin{eqnarray} 
\Hred^{\smallpm}_x &=&(\pm 1 -1) +\sum_{k}^{\calK^{\smallpm}} \Hred^{(k)}_{x} \label{eqn:Hred_x_ksum_abc/pbc} \\ 
\Hred^{\smallpm}_z &=&    \sum_{k}^{\calK^{\smallpm}} \Hred^{(k)}_{z}   \;, \label{eqn:Hred_z_ksum_abc/pbc}
\end{eqnarray}
Here for each $k$-vector we have:
\begin{eqnarray} 
\Hred^{(k)}_{x} &=&-2\PauliTau^z_k=-2\versorz\cdot\bpaulitau_k    \label{eqn:Hred_x_kspin_abc/pbc} \\
\Hred^{(k)}_{z} &=& (2\sin k) \, \PauliTau^x_k - (2\cos k) \, \PauliTau^z_k=-2\versorb_{k}\cdot \bpaulitau_k\;, 
\label{eqn:Hred_z_kspin_abc/pbc}
\end{eqnarray}
where we defined the unit vectors $\versorz=(0,0,1)^T$ and $\versorb_{k}=(-\sin k,0, \cos k)^T$.

\subsection{Jordan-Wigner transformation for the digitized dynamics}
The pseudo-spin representation is useful to discuss the digital dynamics induced by the reduced Hamiltonians introduced in the previous section. 
To simplify the notation, we omit in this section the explicit indication of the boundary conditions used, and the tilde in the reduced spin states. 
In this representation, the initial state $\ket{\psi_0}$, being the ground state of $\Hred_x$, corresponds to a state where all pseudo-spins are aligned 
along the $\versorz$ axis. 
The initial pseudo-spin magnetization $\btau_k(0)$ is therefore 
\begin{equation}
\btau_k(0)=\bra{\psi_0}\bpaulitau_k\ket{\psi_0}=\versorz \;.
\end{equation}
Then, starting from such an initial condition $\btau_k(0)=\versorz$, the $\Hred_z$ and $\Hred_x$ Hamiltonians are used to 
perform a sequence of rotations on the pseudo-spins. 
The action of each digital step is obtained from the identity
\begin{eqnarray}
\nep^{i   \gamma_m \Hred_z}\nep^{i   \beta_m \Hred_x}
\bpaulitau_k 
\nep^{-i \beta_m \Hred_x}\nep^{-i \gamma_m \Hred_z}  
&=& \Rrot_{\versorz}(4 \beta_m )\Rrot_{\bvec_{k}}(4 \gamma_m ) 
\bpaulitau_k \label{eq:discratezRotation}
\end{eqnarray}
where $R_{\hat{\boldsymbol{\omega}}}(\theta)$ is the $3\times3$ matrix associated with a rotation of 
an angle $\theta$ around the unit vector $\hat{\boldsymbol{\omega}}$.
Composing all the rotations appearing in the definition of $\Uevol_{\mgdigital}(\bgamma,\bbeta)$, see Eq.~\eqref{eqn:Udigital} in the main text, 
one gets the final pseudo-spin magnetization $\btau_k(\bgamma,\bbeta)$: 
\begin{eqnarray}
\btau_k(\bgamma,\bbeta)=\bra{\psi_{\Ptrot}(\bgamma,\bbeta)}\bpaulitau_k\ket{\psi_{\Ptrot}(\bgamma,\bbeta)} 
&=& \bra{\psi_0}\Uevol^\dagger_{\mgdigital}(\bgamma,\bbeta) \bpaulitau_k\Uevol_{\mgdigital}(\bgamma,\bbeta)\ket{\psi_0} \nonumber \\
&=&\left( \Tprod{\Ptrot}_{m=1}  \Rrot_{\versorz}(4 \beta_m )\Rrot_{\bvec_k}(4 \gamma_m) \right) \versorz \;.\label{eqn:pseudospin_full_rotation}
\end{eqnarray}
Eq.~\eqref{eqn:pseudospin_full_rotation} holds both with PBC and ABC. 
However, since $\calK^{\smallp}$ and $\calK^{\smallm}$ are not equal, the wave-vectors that contribute to the energy density 
$e_\Ptrot(\bgamma, \bbeta)=E_\Ptrot(\bgamma, \bbeta)/\Nred$ depend on the boundary condition. 
Indeed, using Eqs.~\eqref{eqn:Hred_z_ksum_abc/pbc}, \eqref{eqn:Hred_z_kspin_abc/pbc}, \eqref{eqn:pseudospin_full_rotation}, 
the energy density can be written as
\begin{eqnarray} \label{eqn:energy_density}
e_\Ptrot(\bgamma, \bbeta) = \frac{E_\Ptrot(\bgamma, \bbeta)}{\Nred } &=&\frac{1}{\Nred}\bra{\psi_{\Ptrot}(\bgamma,\bbeta)} \Hred^{\smallpm}_z \ket{\psi_{\Ptrot}(\bgamma,\bbeta)} 
= -\frac{2}{\Nred}\sum_{k}^{\calK^{\smallpm}} \btau_k(\bgamma,\bbeta) \cdot \versorb_k\nonumber\\
&=& -\frac{2|\calK^{\smallpm}|}{\Nred} + \frac{1}{\Nred}\sum_{k}^{\calK^{\smallpm}} \norm{\btau_k(\bgamma,\bbeta) - \versorb_k}^2\;.
\end{eqnarray}
%
%
where in the last step we used that $\versorb_k$ and $\btau_k$ are unit vectors, and denoted by $|\calK^{\smallpm}|$ the number of $k$-vectors
in $\calK^{\smallpm}$. 

We now consider a full chain of $N$ spins with PBC. 
For the reader convenience, we recall that expression for the residual energy
in such a system is (see Eq.~\ref{eqn:e_res} in the main text):
\begin{equation} \label{eqn:e_res_chain} 
\eres_{\Ptrot}(\bgamma, \bbeta) = \frac{E_\Ptrot(\bgamma, \bbeta)-E_{\min}}{E_{\max}-E_{\min}}=  \frac{1}{2}e_\Ptrot(\bgamma, \bbeta)+\frac{1}{2}\;,
\end{equation}
where we have explicitly used that, for the full chain, $E_{\min}=-N$, $E_{\max}=N$ and $e_{\Ptrot}(\bgamma, \bbeta)=\frac{E_{\Ptrot}(\bgamma, \bbeta)}{N}$.
In the main text we argued, following Ref.~\cite{Mbeng_HybridQVS_arXiv2019} where a detailed proof is given, that for 
$2\Ptrot+2\le N$, $\eres_{\Ptrot}(\bgamma, \bbeta)$ can be equivalently computed in a reduced chain of $\Nred=2\Ptrot+2$ spins and that
changing the boundary conditions of the reduced chain does not affect the value of the residual energy.  
%
Using ABC for the reduced chain is indeed convenient in establishing a non-trivial bound for the residual energy, as detailed in Ref.~\cite{Mbeng_HybridQVS_arXiv2019}.
To see such a variational bound within our JW setting, consider choosing ABC. 
Recalling that $2|\calK^{\smallm}|=\Nred-2$, from Eq.~\ref{eqn:energy_density} and Eq.~\ref{eqn:e_res_chain} we conclude that for $2\Ptrot<N$:
\begin{equation}  \label{eqn:eres_pbc=abc}
\eres_{\Ptrot}(\bgamma, \bbeta) \stackrel{\scriptscriptstyle 2\Ptrot< N}{=}  \frac{1}{2\Ptrot+2} + 
\frac{1}{2\Ptrot+2}{\displaystyle \sum_{k}^{\calK^{\smallm}}} \frac{\norm{\btau_k(\bgamma,\bbeta) - \versorb_k}^2}{2} \geq \frac{1}{2\Ptrot +2}\;.
\end{equation}
For $2\Ptrot \ge N$ we {\em must} use PBC~\cite{Mbeng_HybridQVS_arXiv2019}, hence $2|\calK^{\smallp}|=\Nred$, and we get:
\begin{equation}  \label{eqn:eres_pbc_app}
\eres_{\Ptrot}(\bgamma, \bbeta) \stackrel{\scriptscriptstyle 2\Ptrot\ge N}{=}  
\frac{1}{N}{\displaystyle \sum_{k}^{\calK^{\smallp}}} \frac{\norm{\btau_k(\bgamma,\bbeta) - \versorb_k}^2}{2} \geq 0\;.
\end{equation}
These are the same expressions presented in Eqs.~\eqref{eqn:eres_abc} and \eqref{eqn:eres_pbc} of the main text, as one can immediately show that
the expression for $\epsilon_k(\bgamma,\bbeta)$ in Eq.~\eqref{eqn:eresk_geometrical_def} of the main text is indeed:
\begin{equation}
\epsilon_k(\bgamma,\bbeta) 
= 1-  \versorb_k^T \left( \Tprod{\Ptrot}_{m=1}  \Rrot_{\versorz}(4 \beta_m )\Rrot_{\versorb_k}(4 \gamma_m ) \right)  \versorz
= \frac{\norm{\btau_k(\bgamma,\bbeta) - \versorb_k}^2}{2} \;.
\end{equation}
\end{widetext}

\end{document}